\documentclass[twocolumn,english,aps,prl,showpacs,superscriptaddress]{revtex4-1}
\usepackage[T1]{fontenc}
\usepackage[latin9]{inputenc}
\usepackage{color}
\usepackage{units}
\usepackage{amsmath}
\usepackage{amssymb}
\usepackage{graphicx}
\usepackage{wasysym}

\makeatletter

\newcommand*\LyXThinSpace{\,\hspace{0pt}}

\usepackage{babel}
\usepackage{dsfont}

\usepackage{hyperref}
\def\equationautorefname~#1\null{Equation (#1)\null}

\makeatother

\usepackage{babel}
\IfFileExists{lmodern.sty}{\usepackage{lmodern}}{}

\begin{document}

\title{Long-lived entanglement generation of nuclear spins using coherent
light}

\author{Or Katz}
\email[Corresponding author:]{or.katz@weizmann.ac.il}

\affiliation{Department of Physics of Complex Systems, Weizmann Institute of Science,
Rehovot 76100, Israel}

\affiliation{Rafael Ltd, IL-31021 Haifa, Israel}

\author{Roy Shaham}

\affiliation{Department of Physics of Complex Systems, Weizmann Institute of Science,
Rehovot 76100, Israel}

\affiliation{Rafael Ltd, IL-31021 Haifa, Israel}

\author{Eugene S. Polzik}

\affiliation{Niels Bohr Institute, University of Copenhagen, Blegdamsvej 17, DK-2100
Copenhagen, Denmark.}

\author{Ofer Firstenberg}

\affiliation{Department of Physics of Complex Systems, Weizmann Institute of Science,
Rehovot 76100, Israel}
\begin{abstract}
Nuclear spins of noble-gas atoms are exceptionally isolated from the
environment and can maintain their quantum properties for hours at
room temperature. Here we develop a mechanism for entangling two such
distant macroscopic ensembles by using coherent (\emph{i.e.}~classical)
light input. The interaction between the light and the noble-gas spins
in each ensemble is mediated by spin-exchange collisions with alkali-metal
spins, which are only virtually excited. The relevant conditions for
experimental realizations with $^{3}\text{He}$ or $^{129}\text{Xe}$
are outlined.
\end{abstract}
\maketitle
Quantum entanglement describes correlations between distinct quantum
systems and is often used to set borders between the quantum and classical
worlds \cite{Haroche-popular-1998,Quantum-entanglement}. It is a
valuable resource for quantum information and computing \cite{Nielsen-Chuang,computers,Braunstein-CV-INFORMATION,Gisin-Quantum-communication,Kimble-QUantum-internet}
and for metrology beyond the standard quantum limits \cite{Treutlein-RMP-2018,Polzik-magnetometer-2010}.
Generating and maintaining entanglement in matter systems requires
exquisite control and isolation, as achieved in ensembles of alkali-metal
spins \cite{Polzik-2001,Polzik-RMP-2010,Mitchell-2018}, trapped ions
and atoms \cite{Ion-entanglement,atoms-entanglement}, quantum defects
in crystals \cite{NV-C13-entanglement}, and high-quality mechanical
oscillators \cite{mechanical-resonator-entanglement}.

Rare isotopes of noble-gas atoms, such as \textsuperscript{3}He and
\textsuperscript{129}Xe, have nuclei with nonzero spins. These spins
are exceptionally isolated from the environment and can remain coherent
for extremely long times, exceeding tens of hours above room-temperature
\cite{Gemmel-60-hours-coherence-time-He-2010,Walker-RMP-2017}. Accordingly,
the collective nuclear spin of noble-gas ensembles is the longest-living
macroscopic quantum object currently known. Nevertheless, while these
spin ensembles could potentially maintain entanglement for record
times \cite{Firstenberg-Weak-collisions,Sinatra-HE3-memory-2005},
they do not interact with optical photons. This limits their applicability
for optical quantum communication \cite{Polzik-2001,Kuzmich1,Kuzmich2,Kuzmich3,Polzik-two-cell-theory},
or to advanced sensing applications such as hybrid optomechanical-spin
systems, \textit{e.g.,} for gravitational-wave detection \cite{Polzik-optomechanics,Polzik-ligo}.
In 2007, Pinard and coworkers proposed to entangle \textsuperscript{3}He
ensembles using incoherent collisions with metastable $^{3}\text{He}$
atoms and via adiabatic state transfer with nonclassical light in
an optical cavity \cite{Sinatra-squeezing-He3-2007}. This pioneering
and rather challenging proposal was never realized. 

\begin{figure}[tb]
\begin{centering}
\includegraphics[viewport=0bp 0bp 432bp 284bp,clip,width=8.6cm]{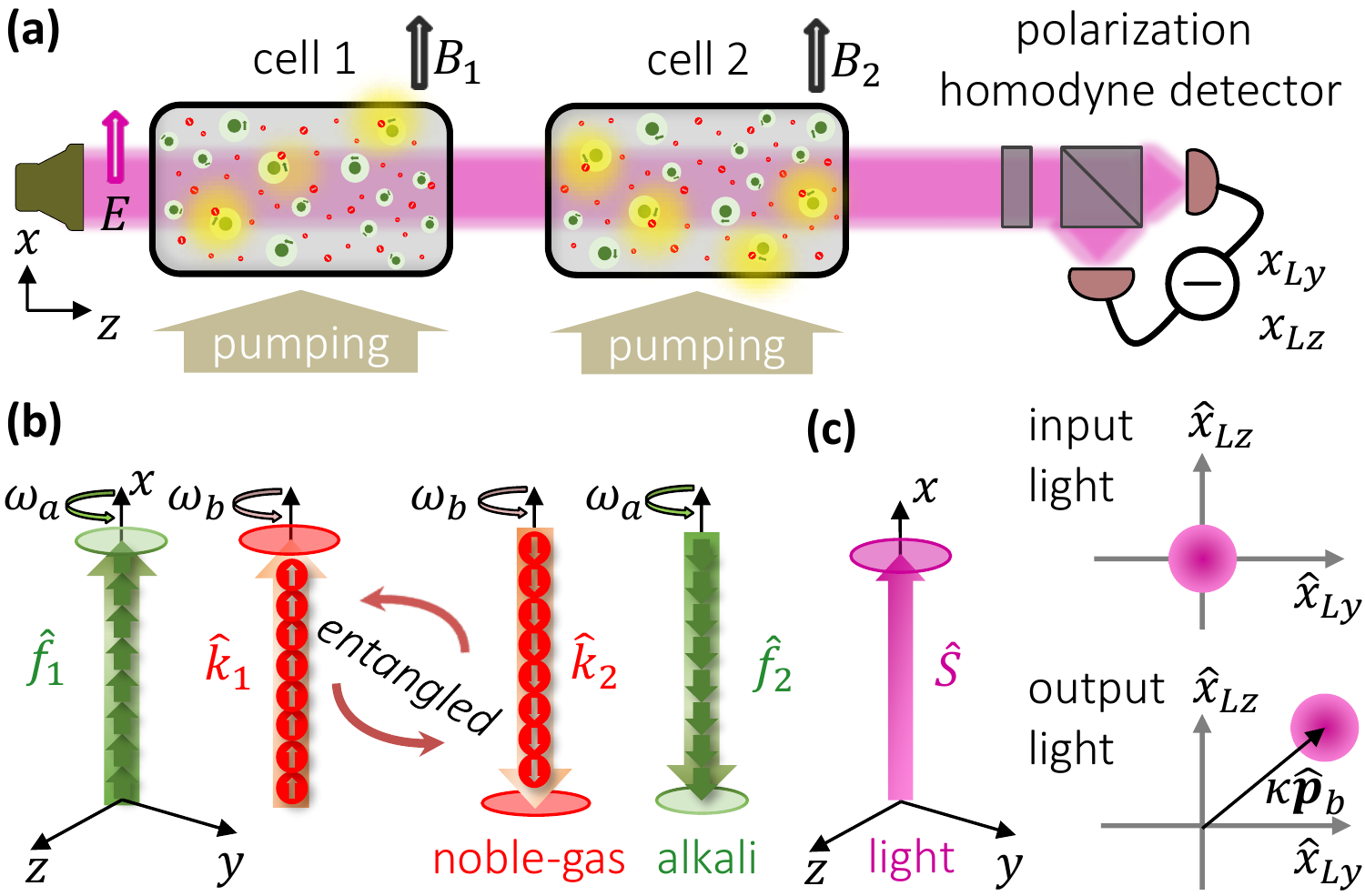}
\par\end{centering}
\centering{}\caption{Entanglement generation of the macroscopic spin-state of two distant
noble-gas ensembles. (a) The physical system consists of two cells
with mixtures of alkali (green) and noble-gas atoms (red). Homodyne
detection of coherent probe light passing through the two cells monitors
the correlated spin precession of the noble-gas ensembles. (b) Collective
spin-states of polarized alkali and noble-gas atoms. The shaded disks
denote quantum spin fluctuations. (c) Polarization state of linearly-polarized
probe, and its rotation via indirect Faraday interaction with the
noble-gas spins, as described by Eq.~(\ref{eq:input-output relations}).
The in-phase $(\hat{x}_{Ly})$ and out-of-phase $(\hat{x}_{Lz})$
components of the probe commute and can be simultaneously measured.
Shaded purple disks denote the photon shot-noise. \label{fig: setup} }
\end{figure}

Here we develop a readily feasible scheme for entangling two macroscopic
ensembles of noble-gas spins contained in distant cells, as shown
in Fig.~\ref{fig: setup}. Our scheme employs the archetypal mechanism
for entanglement of spin ensembles, based on continuous measurement
of spin fluctuations by off-resonant Faraday rotation of probe light
\cite{Polzik-two-cell-theory}. This mechanism was successfully employed
to entangle distant alkali spin ensembles \cite{Polzik-2001}. While
there is no direct interaction between light and noble-gas spins,
we propose to use auxiliary ensembles of alkali-metal atoms as mediators.
The alkali mediators are optically-accessible and couple to the noble-gas
spins via coherent spin-exchange collisions \cite{Firstenberg-Weak-collisions}.
We show that continuous optical measurement of the alkali spins generates
a vital entanglement between the noble-gas ensembles. At the same
time, dissipation and fluctuations of the alkali spins can be circumvented
by introducing a frequency mismatch, such that quantum correlations
are mediated without actual excitations of the (alkali) mediators.
We outline the physical conditions for experiments with \textsuperscript{3}He-K
and \textsuperscript{129}Xe-Rb mixtures towards a demonstration of
long-lived entanglement of macroscopic systems.
\begin{figure*}[t]
\begin{centering}
\includegraphics[viewport=0bp 0bp 709bp 113bp,clip,width=17cm]{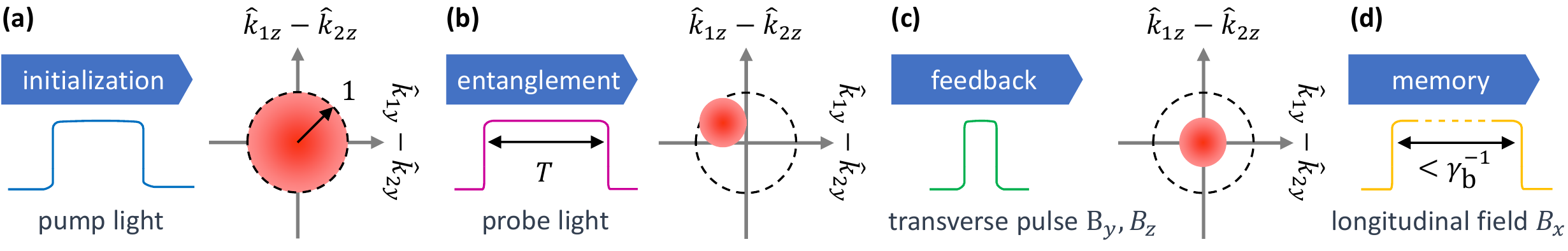}
\par\end{centering}
\centering{}\caption{Sequence for generation and storage of entanglement. (a) The noble-gas
ensembles are pumped to coherent spin states with vacuum fluctuations
of radius $\text{std}(\hat{\mathrm{k}}_{1y}-\hat{\mathrm{k}}_{2y})=\text{std}(\hat{\mathrm{k}}_{1z}-\hat{\mathrm{k}}_{2z})=1$.
Dashed circles mark the entanglement criterion from Eq.~(\ref{eq:Entangelment inequality}).
(b) Homodyne detection of the probe light, via the Faraday interaction
(Fig.~\ref{fig: setup}c), leads to (conditional) squeezing and displacement
of the spin-state. $\hat{\mathrm{k}}_{1y}-\hat{\mathrm{k}}_{2y}$
and $\hat{\mathrm{k}}_{1z}-\hat{\mathrm{k}}_{2z}$ commute, and their
combined uncertainty can be smaller than 1. (c) A short transverse
magnetic-field pulse rotates the spin-state, yielding an unconditioned
entanglement, satisfying inequality~(\ref{eq:Entangelment inequality}).
(d) During the memory time, application of a large magnetic-field
decouples the noble-gas and alkali spins. The memory lifetime is governed
by the long coherence-time of the noble-gas spins. \label{fig: sequence}}
\end{figure*}

Before diving into the detailed model, we consider a simplified picture
of the interaction mechanisms within each cell, presenting the emergence
of the Faraday interaction between light and optically-inaccessible
spins. We describe quantum excitations of the alkali spins by the
bosonic operators $\hat{f},\hat{f}^{\dagger},$ excitations of noble-gas
spins by $\hat{k},\hat{k}^{\dagger},$ and the polarization state
of probe light by the canonical bosonic operators $\hat{x}_{\mathrm{L}}$
and $\hat{p}_{\mathrm{L}}$. The probe couples to the alkali ground-level
spins via the optically-excited levels. These levels are subject to
rapid relaxation at a rate $\Gamma_{e}$ due to spontaneous emission
and buffer-gas broadening, leading to spin relaxation and to probe
attenuation. Detuning the probe by $|\delta_{\mathrm{e}}|\gg\Gamma_{\mathrm{e}}$
from the optical transition circumvents this relaxation, rendering
the atom-photon interaction dispersive. The excited-level spins then
adiabatically follow the ground-level spins, yielding the Faraday
interaction $\mathcal{H}_{\mathrm{L-a}}=i\hbar Q\hat{p}_{\mathrm{L}}(\hat{f}^{\dagger}-\hat{f})/\sqrt{2}$
between the probe and the alkali spins. $\mathcal{H}_{\mathrm{L-a}}$
describes the polarization rotation of far-detuned probe and the resulting
alkali-spin rotation at the rate $Q\propto1/\delta_{\mathrm{e}}$
\cite{Polzik-RMP-2010}.

The coherent coupling of the alkali spins to the noble-gas spins is
described by the exchange Hamiltonian $\mathcal{H}_{\mathrm{a-b}}=\hbar J(\hat{f}^{\dagger}\hat{k}+\hat{k}^{\dagger}\hat{f})$,
where $J$ is the collective exchange-rate due to atomic collisions
\cite{Firstenberg-Weak-collisions}. The resonance conditions for
this coupling are governed by the non-interacting Hamiltonian $\mathcal{H}_{0}=\hbar\omega_{\mathrm{a}}\hat{f}^{\dagger}\hat{f}+\hbar\omega_{\mathrm{b}}\hat{k}^{\dagger}\hat{k}$,
where the difference in precession frequencies $\Delta=\omega_{\mathrm{a}}-\omega_{\mathrm{b}}$
is tunable with an external magnetic field. 

The alkali spins are prone to fast dephasing at a rate $\gamma_{\mathrm{a}}$
due to photon absorption, collisions with different atoms and with
the cell walls. Here again, the detuning $(\Delta)$ determines to
what extent this fast alkali relaxation affects the noble-gas spins.
On resonance $(\left|\Delta\right|\lesssim\gamma_{\mathrm{a}},J)$,
the noble-gas spins inherit the alkali-spin relaxation \cite{Firstenberg-Weak-collisions},
whereas off resonance ($|\Delta|\gg J,\gamma_{\mathrm{a}}$), the
interaction is dispersive, suppressing the relaxation induced by the
alkali by a factor $\gamma_{\mathrm{a}}/\Delta\ll1$. The alkali spins
then adiabatically follow the noble-gas spins, yielding the overall
Hamiltonian $\mathcal{H}_{\mathrm{L-b}}=i\hbar QJ\hat{p}_{\mathrm{L}}(\hat{k}-\hat{k}^{\dagger})/(\sqrt{2}\Delta)$
in a frame rotating at $\omega_{\mathrm{b}}$ when $|\Delta|\gg J,Q,$
up to shifts proportional to $Q^{2}/\Delta$ and $J^{2}/\Delta$.
We thus arrive at an indirect Faraday interaction of light with noble-gas
spins via virtual excitations of alkali spins.

The concept described above can be applied for entangling two distant
noble-gas spin ensembles using probe light and alkali spins {[}Fig.~\ref{fig: setup}(a){]}.
Each cell contains $N_{\mathrm{b}}$ noble-gas atoms with spin-1/2,
initially polarized along the quantization axis $\boldsymbol{e}_{x}$.
Ensemble $i=1,\,(i=2)$ is polarized upwards $+\boldsymbol{e}_{x}$
(downwards $-\boldsymbol{e}_{x}$). Given the spin operators $\hat{\boldsymbol{\mathrm{k}}}_{i}^{(n)}$
of the \textit{n}-th noble-gas atom in the \textit{i}-th cell, we
define the normalized macroscopic spin operator $\hat{\boldsymbol{\mathrm{k}}}_{i}\equiv M_{\mathrm{b}}^{-1/2}\sum_{n=1}^{N_{\mathrm{b}}}\hat{\boldsymbol{\mathrm{k}}}_{i}^{\left(n\right)}$for
each ensemble. The total magnetization $M_{\mathrm{b}}=P_{\mathrm{b}}N_{\mathrm{b}}/2$
depends on the initial degree of polarization $P_{\mathrm{b}}\le1$.
For $M_{\mathrm{b}}\gg1$ and fully polarized ensembles $(P_{\mathrm{b}}=1)$,
the initial states are known as coherent spin-states (CSS). A partially
polarized ensemble of spin-1/2 atoms may be seen as a mixture of $P_{\mathrm{b}}N_{\mathrm{b}}$
polarized atoms and $(1-P_{\mathrm{b}})N_{\mathrm{b}}$ unpolarized
atoms, only reducing the coherent interaction strength \cite{Polzik-RMP-2010}.
The two ensembles have definitive collective spin along $\boldsymbol{e}_{x}$
with a classical measurement outcome $\langle\hat{\mathrm{k}}_{ix}\rangle=\pm M_{\mathrm{b}}^{1/2}$
and negligible variance, where henceforth the symbol `$\pm$' stands
for `+' in cell $i=1$, and for `$-$' in cell $i=2$. On the other
hand, the transverse components of the normalized collective spin
$\hat{\mathrm{k}}_{iy}$ and $\hat{\mathrm{k}}_{iz}$ satisfy the
commutation relation $[\hat{\mathrm{k}}_{iy},\hat{\mathrm{k}}_{jz}]=\pm i\delta_{ij}$
and consequently are governed by quantum fluctuations. These operators
are normalized and unitless, giving the collective spin variance in
units of vacuum noise. These fluctuations, known as atom-projection
noise, are zero on average and have a nonzero variance, satisfying
the Robertson inequality $4\text{var}(\hat{\mathrm{k}}_{iy})\text{var}(\hat{\mathrm{k}}_{iz})\geq|\bigl\langle[\hat{\mathrm{k}}_{iy},\hat{\mathrm{k}}_{iz}]\bigr\rangle|^{2}=1$,
where $\text{var}(\hat{\mathrm{k}}_{iy})=\text{var}(\hat{\mathrm{k}}_{iz})$
for CSS. Visually, these fluctuations can be represented as a small
uncertainty disk around the classical spin vector, as shown in Fig.~\ref{fig: setup}(b). 

Two spin ensembles are entangled if their quantum fluctuations are
correlated, as in a two-mode squeezed state. For spins of equal magnitude
$|\langle\hat{\mathrm{k}}_{1x}\rangle|=|\langle\hat{\mathrm{k}}_{2x}\rangle|$,
a sufficient criterion for EPR-type entanglement is given by \cite{Polzik-2001,Duan-2000}
\begin{equation}
\text{var}(\hat{\mathrm{k}}_{1y}-\hat{\mathrm{k}}_{2y})+\text{var}(\hat{\mathrm{k}}_{1z}-\hat{\mathrm{k}}_{2z})<2.\label{eq:Entangelment inequality}
\end{equation}
Therefore, simultaneous measurement of the nonlocal observables $\hat{\mathrm{k}}_{1y}-\hat{\mathrm{k}}_{2y}$
and $\hat{\mathrm{k}}_{1z}-\hat{\mathrm{k}}_{2z}$ generates entanglement,
if the total noise variance of the two cells is less than two vacuum-noise
units. Such measurement is allowed for oppositely oriented spins $\langle\hat{\mathrm{k}}_{1x}\rangle=-\langle\hat{\mathrm{k}}_{2x}\rangle$,
for which $\hat{\mathrm{k}}_{1y}-\hat{\mathrm{k}}_{2y}$ and $\hat{\mathrm{k}}_{1z}-\hat{\mathrm{k}}_{2z}$
commute.

We measure the noble-gas spins using alkali spins and a probe field.
Each cell contains $N_{\mathrm{a}}$ alkali atoms, polarized to a
polarization degree $P_{\mathrm{a}}\le1$ (using auxiliary circularly-polarized
pump beams) along the same directions $\pm\boldsymbol{e}_{x}$ as
the noble-gas spins. We define for each cell the normalized macroscopic
alkali-spin operator $\boldsymbol{\hat{\boldsymbol{\mathrm{f}}}}_{i}\equiv M_{\mathrm{a}}^{-1/2}\sum_{m=1}^{N_{\mathrm{a}}}\boldsymbol{\hat{\boldsymbol{\mathrm{f}}}}_{i}^{(m)}$,
where $M_{\mathrm{a}}=P_{\mathrm{a}}N_{\mathrm{a}}(I+1/2)$ is the
alkali magnetization, and $I$ is the alkali nuclear spin. Similarly
to the noble-gas spins, $\hat{\mathrm{f}}_{ix}$ are considered classical,
with $\langle\hat{\mathrm{f}}_{ix}\rangle=\pm M_{\mathrm{a}}^{1/2}$,
whereas $\hat{\mathrm{f}}_{iy}$ and $\hat{\mathrm{f}}_{iz}$ are
governed by quantum fluctuations. The probe is a square pulse of duration
$T$, propagating along $\boldsymbol{e}_{z}$ with initial linear
polarization $\boldsymbol{e}_{x}$. We represent its state by the
normalized Stokes operators $\hat{\boldsymbol{S}}(z)$ where $\langle\hat{S}_{x}\rangle^{2}=M_{\mathrm{L}}$
is the total number of photons in the pulse, and $\hat{S}_{y},\hat{S}_{z},$
describe the ellipticity of the polarization-state subject to quantum
polarization-fluctuations.

The Hamiltonian describing the interactions in the system is given
by \cite{Polzik-2001,Firstenberg-Weak-collisions} 
\begin{equation}
\mathcal{V}=\hbar J\bigl(\boldsymbol{\hat{\mathrm{f}}}_{1}\cdot\hat{\boldsymbol{\mathrm{k}}}_{1}+\boldsymbol{\hat{\mathrm{f}}}_{2}\cdot\hat{\boldsymbol{\mathrm{k}}}_{2}\bigr)+\hbar Q\bigl(\hat{\mathrm{f}}_{1z}+\hat{\mathrm{f}}_{2z}\bigr)\int\frac{dz'}{L}\hat{S}_{z}(z').\label{eq:Interaction Hamiltonian}
\end{equation}
The first term describes a mutual precession of the alkali and noble-gas
spins around each other at a rate $J$. It manifests the coherent
collective coupling between these spins via multiple weak spin-exchange
collisions \cite{Firstenberg-Weak-collisions}. The second term in
Eq.~(\ref{eq:Interaction Hamiltonian}) describes the dispersive
interaction of the alkali spins with the far-detuned probe traversing
the two cells \cite{Polzik-RMP-2010}. The spin components along the
optical axis $(\hat{\mathrm{f}}_{1z}+\hat{\mathrm{f}}_{2z})$ govern
the Faraday rotation of the light polarization, while circularly-polarized
light $(\hat{S}_{z})$ acts back to rotate the spins via light-shifts.
The coupling rate is given by $Q=$$(a/T)\sqrt{M_{\mathrm{a}}M_{\mathrm{L}}}$,
where $a\propto1/\delta_{\mathrm{e}}$ is the unitless optical-coupling
coefficient \cite{Polzik-RMP-2010,Romalis-stroboscopic-2011} and
$L$ is the length of each cell. See Supplementary Material for detailed
expressions of $J,Q,$ and $a$ \cite{SM}. 

To generate entanglement, we set common precession frequencies $(\omega_{\text{a}},\omega_{\text{b}})$
in the two cells, by tuning the magnetic fields and the light-shifts
induced by the pumps in each cell \cite{SM}. We describe the spin
dynamics in a common rotating frame, defined by $\hat{\boldsymbol{\mathrm{k}}}_{i}\rightarrow R_{x}\bigl(\omega_{\mathrm{b}}t\bigr)\hat{\boldsymbol{\mathrm{k}}}_{i}$
and $\hat{\boldsymbol{\mathrm{f}}}_{i}\rightarrow R_{x}\bigl(\omega_{\mathrm{b}}t\bigr)\hat{\boldsymbol{\mathrm{f}}}_{i}$,
where $R_{x}\left(\theta\right)$ rotates a vector by an angle $\theta$
around $\boldsymbol{e}_{x}$. In this frame, the alkali spins precess
at frequency $\Delta=\omega_{\mathrm{a}}-\omega_{\mathrm{b}}.$ 

We now take the off-resonance regime $\Delta\gg\gamma_{\mathrm{a}},J,Q$
and first present the results for negligible relaxations. Given the
interaction Hamiltonian (\ref{eq:Interaction Hamiltonian}), we find
that the transverse fluctuations $\hat{\mathrm{f}}_{iy},\,\hat{\mathrm{f}}_{iz}$
of the alkali spins adiabatically follow the noble-gas spins-fluctuations,
and the probe polarization,
\begin{align}
\boldsymbol{\hat{\mathrm{f}}}_{i} & =\pm\frac{J}{\Delta}\hat{\boldsymbol{\mathrm{k}}}_{i}\pm\frac{Q}{\Delta}\hat{S}_{z}\boldsymbol{e}(t),\label{eq:adiabatic J}
\end{align}
where $\boldsymbol{e}(t)=\sin(\omega_{\mathrm{b}}t)\boldsymbol{e_{y}}+\cos(\omega_{\mathrm{b}}t)\boldsymbol{e_{z}}$
is the optical axis in the rotating frame. Thus, the large frequency
mismatch $\Delta$ renders the interaction dispersive, moderating
the response of the alkali spins to both spin-exchange and back-action
of light.

We use Eqs.~(\ref{eq:Interaction Hamiltonian}-\ref{eq:adiabatic J})
to derive the Heisenberg-Langevin equations for the transverse operators
$\hat{S}$ and $\hat{\boldsymbol{\mathrm{k}}}_{i}$ \cite{SM}. First,
we find that the difference between the noble-gas spins remains constant
\begin{align}
\partial_{t}\bigl(\hat{\boldsymbol{\mathrm{k}}}_{1}-\hat{\boldsymbol{\mathrm{k}}}_{2}\bigr) & =0.\label{eq:QND interaction}
\end{align}
Importantly, the preparation of the two cells with oppositely oriented
spins eliminates the back-action effect {[}second term in Eq.~(\ref{eq:adiabatic J}){]}
of the probe on the operator $\hat{\boldsymbol{\mathrm{k}}}_{1}-\hat{\boldsymbol{\mathrm{k}}}_{2}$.
Second, we find that $\hat{\boldsymbol{\mathrm{k}}}_{1}-\hat{\boldsymbol{\mathrm{k}}}_{2}$
determines the evolution of the probe polarization along the cell
\begin{equation}
\partial_{z}\hat{S}_{y}=\frac{QJT}{L\Delta}\bigl(\hat{\boldsymbol{\mathrm{k}}}_{1}-\hat{\boldsymbol{\mathrm{k}}}_{2}\bigr)\cdot\boldsymbol{e}(t).\label{eq:homodyne}
\end{equation}
Equation (\ref{eq:homodyne}) manifests the indirect Faraday interaction
between the probe and the noble-gas spins, with the outgoing polarization
$\hat{S}_{y}(L)$ providing a monitor of $\hat{\boldsymbol{\mathrm{k}}}_{1}-\hat{\boldsymbol{\mathrm{k}}}_{2}$.
In particular, a simultaneous measurement of the in-phase and out-of-phase
components of $\hat{S}_{y}(L)$ via homodyne detection yields the
nonlocal spin components $\hat{\mathrm{k}}_{1y}-\hat{\mathrm{k}}_{2y}$
and $\hat{\mathrm{k}}_{1z}-\hat{\mathrm{k}}_{2z}$, respectively. 

The procedure for entanglement generation is shown in Fig.~\ref{fig: sequence}.
Initially, homodyne measurement of the probe, which underwent the
evolution in Eq.~(\ref{eq:homodyne}), drives the noble-gas ensembles
to a nonclassical two-mode squeezed state, displaced according to
the measurement outcome \cite{Polzik-RMP-2010}. Subsequently, feeding-back
the measurement outcome to rotate the spins (using a short magnetic
pulse) sets the mean value of their squeezed components to zero, yielding
unconditioned entanglement. 
\begin{figure}[t]
\begin{centering}
\includegraphics[viewport=0bp 0bp 663bp 283bp,clip,width=8.6cm]{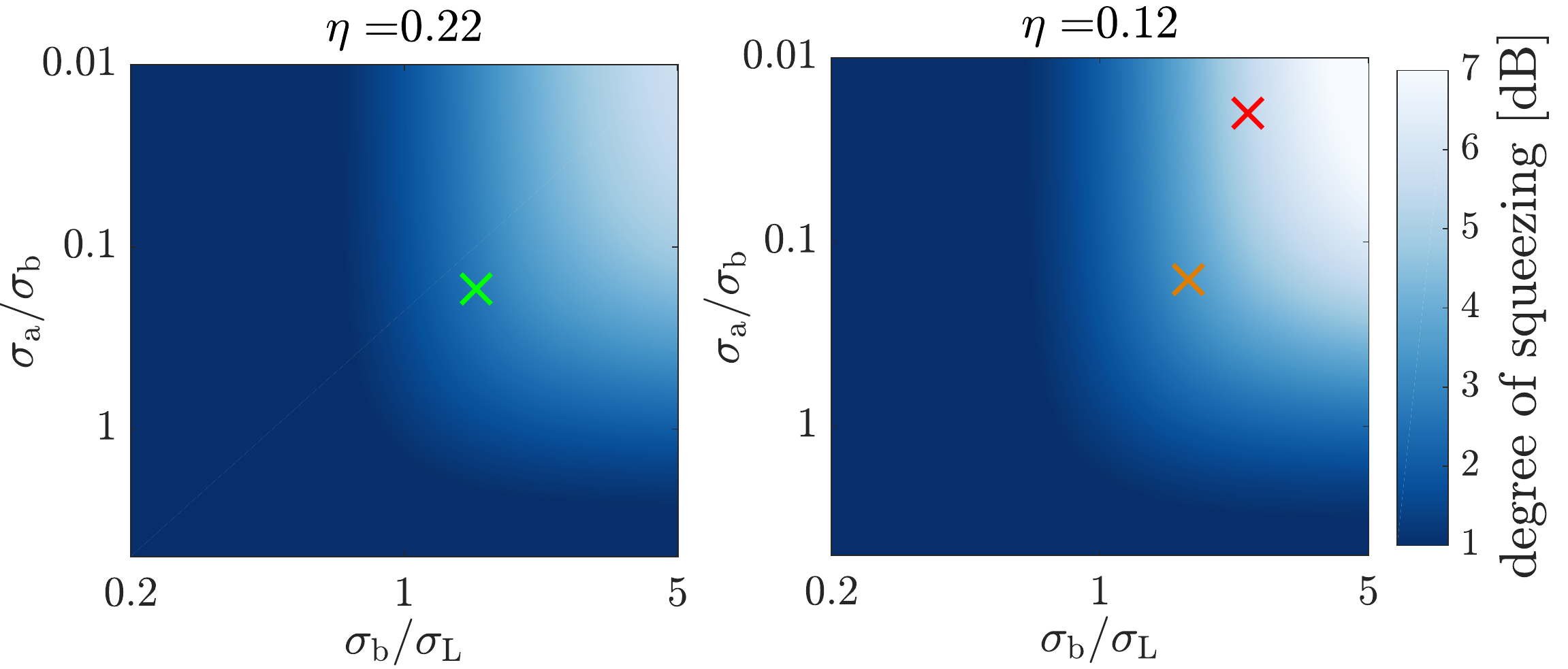}
\par\end{centering}
\centering{}\caption{Attainable degree of two-mode spin squeezing for noble-gas ensembles.
We present results for both $\eta=0.22$ and $\eta=0.12$, where $\eta$
characterizes the fractional decoherence of the noble-gas spins during
the entangling process. The parameters $\sigma_{\mathrm{a}},\,\sigma_{\mathrm{b}},$
and $\sigma_{\mathrm{L}}$ denote the contributions of the alkali
spin-projection noise, noble-gas spin-projection noise, and photon
shot-noise, respectively, to the optical measurements. The squeezing
is maximized when the noble-gas noise $\sigma_{\mathrm{b}}$ dominates
the measurement. The calculations are done using Eq.~(\ref{eq:chi}),
with $\sigma_{\mathrm{b}}/\sigma_{\mathrm{L}}=\kappa\sqrt{1-\epsilon}$
and $\sigma_{\mathrm{a}}/\sigma_{\mathrm{b}}=\varrho$. The crosses
mark proposed working points with \textsuperscript{129}Xe-\textsuperscript{87}Rb
(green) and \textsuperscript{3}He-K (red,orange).\label{fig: squeezing map} }
\end{figure}

To quantify this process, we define canonical operators for the probe
$\boldsymbol{\hat{x}}_{\text{L}}\left(z\right)=\sqrt{2}\int_{0}^{T}\hat{S}_{y}\left(z\right)\boldsymbol{e}\left(t\right)dt/T$
and $\boldsymbol{\hat{p}}_{\text{L}}\left(z\right)=\sqrt{2}\int_{0}^{T}\hat{S}_{z}\left(z\right)\boldsymbol{e}\left(t\right)dt/T,$
and nonlocal canonical operators for the noble-gas spins $\boldsymbol{\hat{x}}_{\text{b}}\left(t\right)=\boldsymbol{e_{x}}\times(\hat{\boldsymbol{\mathrm{k}}}_{1}+\hat{\boldsymbol{\mathrm{k}}}_{2})/\sqrt{2}$
and $\boldsymbol{\hat{p}}_{\text{b}}\left(t\right)=(\hat{\boldsymbol{\mathrm{k}}}_{1}-\hat{\boldsymbol{\mathrm{k}}}_{2})/\sqrt{2}$.
These constitute two independent Harmonic oscillators. The total evolution
is then given by a set of input-output relations, obtained by integration
of Eqs.~(\ref{eq:QND interaction}-\ref{eq:homodyne}) \cite{SM}
\begin{equation}
\begin{aligned}\boldsymbol{\hat{x}}_{\text{L}}^{\text{out}} & =\boldsymbol{\hat{x}}_{\text{L}}^{\text{in}}+\kappa\boldsymbol{\hat{p}}_{\text{b}}^{\text{in}}\\
\boldsymbol{\hat{p}}_{\text{b}}^{\text{out}} & =\boldsymbol{\hat{p}}_{\text{b}}^{\text{in}}.
\end{aligned}
\label{eq:input-output relations}
\end{equation}
The input components of the probe $\boldsymbol{\hat{x}}_{\text{L}}^{\text{in}},\,\boldsymbol{\hat{p}}_{\text{L}}^{\text{in}}$
comprise the photon shot-noise at $z=0$, and the output components
$\boldsymbol{\hat{x}}_{\text{L}}^{\text{out}},\,\boldsymbol{\hat{p}}_{\text{L}}^{\text{out}}$
describe the probe state at $z=2L$ after the cells. Similarly, the
noble-gas spin operators $\boldsymbol{\hat{x}}_{\text{b}}^{\text{in}},\,\boldsymbol{\hat{p}}_{\text{b}}^{\text{in}}$
comprise the atomic projection-noise at $t=0$, and $\boldsymbol{\hat{x}}_{\text{b}}^{\text{out}},\,\boldsymbol{\hat{p}}_{\text{b}}^{\text{out}}$
describe the collective spin-state at $t=T$. Therefore, Eqs.~(\ref{eq:input-output relations})
describe the Faraday rotation $(\boldsymbol{\hat{x}}_{\text{L}}^{\text{out}})$
of the linearly polarized input light $(\boldsymbol{\hat{x}}_{\text{L}}^{\text{in}})$
by the total noble-gas spin $(\boldsymbol{\hat{p}}_{\text{b}}^{\text{in}})$,
as shown in Fig.~\ref{fig: setup}(c), with no back-action $(\boldsymbol{\hat{p}}_{\text{b}}^{\text{out}}=\boldsymbol{\hat{p}}_{\text{b}}^{\text{in}})$.
The unitless coupling constant $\kappa\equiv QJT/\Delta$ quantifies
the net polarization rotation of the probe. It characterizes the measurement
strength of the noble-gas spins with respect to the photon shot-noise,
depending on the resonant optical-depth of the alkali ensembles \cite{SM}.

For coherent light and coherent spin-states, the input uncertainties
are at the classical minimum, satisfying $\text{var}(\hat{x}_{\text{L},\alpha}^{\text{in}})=\text{var}(\hat{p}_{\text{L},\alpha}^{\text{in}})=\text{var}(\hat{x}_{\text{b},\alpha}^{\text{in}})=\text{var(}\hat{p}_{\text{b},\alpha}^{\text{in}})=1/2$
with $\alpha=y,z$. Following the measurement, a magnetic pulse feedback
is used for rotating the noble-gas spins from $\boldsymbol{\hat{p}}_{\text{b}}^{\text{out}}=\boldsymbol{\hat{p}}_{\text{b}}^{\text{in}}$
to $\boldsymbol{\hat{p}}_{\text{b}}^{\text{in}}+G\boldsymbol{\hat{x}}_{\text{L}}^{\text{out}}$.
The feedback proportionality constant $G$ can be optimally chosen
to minimize $\text{var}(\hat{p}_{\text{b},\alpha}^{\text{out}})=(2+2\kappa^{2})^{-1}$
for both $\alpha=y,z$. Identifying $\text{var(}\hat{p}_{\text{b},\alpha}^{\text{out}})=\exp(-2\xi)/2$
as the degree of two-mode squeezing, we obtain the squeezing parameter
$\xi=\ln(1+\kappa^{2})/2$. Evidently, any system with $\kappa>0$
yields nonzero squeezing and satisfies the inequalities $\text{var}\bigl(\hat{p}_{\text{b},\alpha}^{\text{out}}\bigr)<1/2$,
thus satisfying the entanglement condition in Eq.~(\ref{eq:Entangelment inequality}).
We therefore conclude that our scheme correlates the spin-states of
two distant noble-gas ensembles, generating unconditional entanglement. 

We now return to consider relaxation processes expected in realistic
conditions. The mechanisms dominating the relaxation rate $\gamma_{\mathrm{sd}}$
of the alkali spin are absorption of probe photons, collisions with
noble-gas atoms, spin destruction during alkali collisions, and collisions
with the cell walls \cite{Firstenberg-Weak-collisions,Happer-Book,firstenberg-Hybrid_spin_exchange-2015,Katz-storage-of-light-2018}.
Continuous optical-pumping at a rate $R_{\text{op}}$ can be used
to maintain a constant alkali magnetization $M_{\mathrm{a}}=P_{\mathrm{a}}N_{\mathrm{a}}(I+1/2),$
with $P_{\mathrm{a}}=R_{\text{op}}/\gamma_{\mathrm{a}}$ and $\gamma_{\mathrm{a}}=\gamma_{\text{sd}}+R_{\text{op}}$.
The noble gas is hyperpolarized via spin-exchange optical-pumping
(SEOP) at a high magnetic field prior to the experiment \cite{Happer-Book,Happer-1984}.
For polarized alkali spins, the decoherence rate of the noble-gas
spins is $\Gamma_{\mathrm{b}}=\gamma_{\mathrm{b}}+(J/\Delta)^{2}\gamma_{\mathrm{a}}$;
it inherits a fraction $(J/\Delta)^{2}$ of the alkali decoherence
rate $\gamma_{\mathrm{a}}$, which often dominates $\Gamma_{\mathrm{b}}$
\cite{Romalis-2002}. At low alkali densities, $\gamma_{\mathrm{b}}$
is typically limited by technical magnetic inhomogeneities to $\gamma_{\mathrm{b}}\lesssim(\mathrm{minute})^{-1}$
for \textsuperscript{129}Xe and $\gamma_{\mathrm{b}}\lesssim(\mathrm{hour})^{-1}$
for \textsuperscript{3}He \cite{Gemmel-60-hours-coherence-time-He-2010,Walker-RMP-2017,Happer-1985}.

These relaxation processes are accompanied by noise, which increases
the measurement variance and limits $\xi$. We generalize Eqs.~(\ref{eq:input-output relations})
and include the relaxation and noise effects, deriving the best attainable
two-mode squeezing parameter \cite{SM}
\begin{equation}
\xi=\frac{1}{2}\ln\left(\frac{\kappa^{2}\left(1-\epsilon\right)\left(1+\varrho\right)+1}{\kappa^{2}\left(1-\epsilon\right)\left(\eta+\varrho\right)+1}\right).\label{eq:chi}
\end{equation}
Here $\epsilon=4\gamma_{\text{L}}L$ denotes the total fraction of
scattered probe photons, $\eta=2\Gamma_{\mathrm{b}}T$ denotes the
fraction of decohered noble-gas spins, and $\varrho=4q\gamma_{\mathrm{a}}/(J^{2}T)$
characterizes the ratio between the contributions of alkali spins
and noble-gas spins to the projection noise. The unitless parameter
$q(I,P_{\mathrm{a}})\geq1$ quantifies the increase of alkali projection-noise
(variance) due to imperfect spin-polarization, where $q(0,P_{a})=q(I,1)=1$
\cite{Romalis-stroboscopic-2011}. Equation (\ref{eq:chi}) guarantees
the generation of entanglement between the two ensembles for $\eta\ll1$.
Notably, it has the same form as for squeezing two alkali ensembles
\cite{Polzik-RMP-2010} except for the additional parameter $\varrho$.
In Fig.~\ref{fig: squeezing map}, we use Eq.~(\ref{eq:chi}) to
plot the degree of squeezing $\exp(-2\xi)$ of the two noble-gas spin-ensembles
as a function of $\kappa\sqrt{1-\epsilon}$ and $\varrho$ for two
values of $\eta$.

Our entanglement generation scheme can be realized with various alkali
and noble-gas mixtures within a large range of experimental parameters.
Here we present a representative configuration for entangling two
\textsuperscript{3}He ensembles in two cylindrical cells of length
$L=5\,\text{cm}$ and cross-section $A=2\,\text{mm}^{2}$. We consider
a gaseous mixture of 880~Torr \textsuperscript{3}He, 70~Torr N\textsubscript{2},
and a droplet of K at $250^{\circ}\text{C}$. Here $R_{\text{op}}=1.6\gamma_{\mathrm{a}}$
yields $P_{\mathrm{a}}=0.62$ {[}with $q(3/2,P_{\mathrm{a}})=1.22]$
and $P_{\mathrm{b}}=0.56$, assuming $\gamma_{\mathrm{b}}^{-1}=50\,\text{hour}$.
The 400-mW probe is detuned 3 THz from the optical line, and $B_{1}\approx10\,\text{mG}$.
Homodyne detection for $T=200\,\text{msec}$ yields $\kappa=2,\,\epsilon=0.3,\,\eta=0.125,$
and $\varrho=0.162,$ generating 4~dB of two-mode squeezing $(\xi=0.45)$,
which could live for tens of hours. The performance for this configuration
is marked in Fig.~\ref{fig: squeezing map} (orange cross). Other
exemplary experimental configurations, marked in Fig.~\ref{fig: squeezing map}
and detailed in \cite{SM}, yield 6~dB of squeezing for \textsuperscript{3}He-K
mixture (red cross) and 3~dB of squeezing for \textsuperscript{129}Xe-\textsuperscript{87}Rb
mixture (green cross).

The long coherence time within each noble-gas spin ensemble ideally
also applies to the entanglement lifetime, even though each ensemble
comprises a macroscopic number of spins. In the Holstein-Primakoff
approximation, the number of spin excitations is independent of the
total number of spins. Indeed we show in \cite{SM} that the squeezed
quadrature, $\text{var(}\boldsymbol{\hat{p}}_{\text{b}}^{\text{out}})<1/2$,
decays at a constant rate $2\Gamma_{\text{b}}$. 

The long-lived entanglement can be verified by applying an off-resonant
probe pulse, measuring the two spin-ensembles simultaneously by utilizing
the same experimental configuration used for their generation \cite{Polzik-2001}.
Alternatively, the spin of each cell could be measured independently,
and their cross-correlations can be found. In systems featuring strong-coupling
between the alkali and noble-gas $(J\gg\gamma_{\mathrm{a}})$.Transfer times
$J^{-1}$ of a few milliseconds are possible \cite{Firstenberg-Weak-collisions},
realizing fast operations yet maintaining long coherence-times. The
alkali squeezed-state could then be projected using a short probe
pulse.

In summary, we presented a scheme for entangling the collective nuclear
spins of two macroscopic noble-gas ensembles, relying on alkali spin
for obtaining an indirect Faraday interaction between the noble-gas
and light. The role of relaxations has been considered, revealing
that sizable degree of entanglement can be generated at standard experimental
conditions, and maintained for extremely long times. With technologically
available miniature cells \cite{Romalis-stemless-cells,Walker-NMRG,Walker-2019}
and exceptionally long coherence-times, entanglement of hot spin ensembles
holds a promise for realizing new quantum-optics applications and
enhanced sensing at ambient conditions. The scheme could potentially
be extended to generate entanglement in other physical systems having
hybrid electronic and optically-inaccessible nuclear spins, including
quantum dots, diamond color-centers, and rare-earth impurities interacting
with nearby nuclear spins in the crystal.
\begin{acknowledgments}
We acknowledge financial support by the European Research Council
starting investigator grant Q-PHOTONICS 678674, ERC Advanced Grant
QUANTUM-N, the Israel Science Foundation and ICORE, the Pazy Foundation,
the Minerva Foundation with funding from the Federal German Ministry
for Education and Research, and the Laboratory in Memory of Leon and
Blacky Broder. ESP was supported by VILLUM FONDEN under a Villum Investigator Grant, grant no. 25880
\end{acknowledgments}

\onecolumngrid \appendix 
\setcounter{equation}{0}
\setcounter{figure}{0}
\setcounter{table}{0}
\setcounter{page}{1}
\makeatletter
\renewcommand{\theequation}{S\arabic{equation}}
\renewcommand{\thefigure}{S\arabic{figure}}
\renewcommand{\bibnumfmt}[1]{[S#1]}
\renewcommand{\citenumfont}[1]{S#1} 

\part*{Supplementary Information for ``Long-lived entanglement generation
of nuclear spins using coherent light''}

\section{Derivation of the equations of motion}

To derive the equations of motion of the spin and light operators,
we consider the total atomic Hamiltonian $H_{\text{tot}}=\mathcal{H}_{1}+\mathcal{H}_{2}+\mathcal{V}$,
composed of the interaction Hamiltonian $\mathcal{V}$ {[}given in
Eq.~(2) in the main text{]} and the bare atomic Hamiltonians of the
two cells $\mathcal{H}_{1}$ and $\mathcal{H}_{2}$. The bare Hamiltonians
depend on the hyperfine interaction within each alkali atom and on
the energy splitting of the spin levels. They are given by ($i=1,2)$

\begin{equation}
\mathcal{H}_{i}=\sum_{n=1}^{N_{\mathrm{a}}}\left(A_{\text{hpf}}\boldsymbol{\hat{i}}_{i}^{\left(n\right)}\cdot\hat{\boldsymbol{s}}_{i}^{\left(n\right)}+\hbar\tilde{\omega}_{i\mathrm{a}}\hat{s}_{ix}^{\left(n\right)}\right)+\sum_{n=1}^{N_{\mathrm{b}}}\hbar\tilde{\omega}_{i\mathrm{b}}\hat{k}_{ix}^{\left(n\right)},
\end{equation}
In both cells ($i=1,2$), each alkali atom denoted by $m$ has a spin
$\boldsymbol{\hat{\boldsymbol{\mathrm{f}}}}_{i}^{(m)}=\hat{\mathrm{\boldsymbol{s}}}_{i}^{(m)}+\hat{\boldsymbol{\mathrm{i}}}_{i}^{(m)}$,
composed of an electronic spin-1/2 operator $\hat{\mathrm{\boldsymbol{s}}}_{i}^{(m)}$
and a nuclear spin $I>0$ operator $\hat{\boldsymbol{\mathrm{i}}}_{i}^{(m)}$,
coupled by the strong hyperfine interaction \cite{Happer-Book}. $A_{\text{hpf}}$
denotes the hyperfine coupling constant, $\tilde{\omega}_{i\mathrm{a}}=g_{a}B_{i}+\Omega_{i}$
are the energy splittings of the alkali spins and $\tilde{\omega}_{i\mathrm{b}}=g_{\mathrm{b}}B_{i}$
are the level splittings of the noble-gas spins. Here $g_{\mathrm{a}}$
and $g_{\mathrm{b}}$ denote the gyromagnetic ratios of the polarized
alkali and noble-gas spins respectively, where the gyromagnetic ratio
of the alkali spins $g_{\mathrm{a}}$ is $100-1000$ times larger
than that of the noble-gas spins $g_{\mathrm{b}}$. $B_{i}\boldsymbol{e}_{x}$
are the applied magnetic fields in both cells (introducing the Larmor-precession
rates $g_{\mathrm{a}}B_{i}$ and $g_{\mathrm{b}}B_{i}$) and $\Omega_{i}$
denote light shifts induced by the pump beams on the alkali spins.
The hyperfine interaction is the dominant interaction in the Hamiltonian,
and therefore for time scales longer than $1/A_{\text{hpf}}$ and
high polarization of the alkali spin ensemble, we can identify $\hat{\boldsymbol{s}}^{\left(n\right)}\equiv\left[I\right]^{-1}\boldsymbol{\hat{\mathrm{f}}}^{\left(n\right)}$,
where $\left[I\right]=2I+1$.

For alkali-noble-gas mixtures polarized along $\pm\boldsymbol{e}_{x}$,
the collisional interaction leads to coherent exchange between the
quantum spin fluctuations (the transverse spin components), as well
as to fictitious magnetic fields along $\pm\boldsymbol{e}_{x}$ imposed
by each species on the other. Consequently, the total precession frequencies
of the fluctuations of the alkali and noble-gas spins are given respectively
by $\omega_{i\mathrm{a}}=\tilde{\omega}_{i\mathrm{a}}\pm J\sqrt{M_{\mathrm{b}}/M_{\mathrm{a}}}$
and $\omega_{i\mathrm{b}}=\tilde{\omega}_{i\mathrm{b}}\pm J\sqrt{M_{\mathrm{a}}/M_{\mathrm{b}}}$.
To synchronize the precession frequencies in both cells, we set $B_{2}$
and ($\Omega_{2}-\Omega_{1}$) to satisfy $\omega_{1\mathrm{a}}=\omega_{2\mathrm{a}}\equiv\omega_{\mathrm{a}}$
and $\omega_{1\text{b}}=\omega_{2\text{b}}\equiv\omega_{\mathrm{b}}$,
for any choice of $B_{1}$. Under these conditions, the quantum dynamics
of the system derived from the Hamiltonian $H_{\text{tot}}$ is described
by the Heisenberg-Langevin equations for the transverse operators
\begin{align}
\partial_{z}\boldsymbol{\hat{S}}= & \frac{TQ}{L}\bigl(\hat{\mathrm{f}}_{1z}+\hat{\mathrm{f}}_{2z}\bigr)\boldsymbol{e}_{y}-\gamma_{\text{L}}\boldsymbol{\hat{S}}+\boldsymbol{\hat{F}}_{\mathrm{L}}\label{eq: S stokes dynamics-1}\\
\partial_{t}\boldsymbol{\hat{\mathrm{f}}}_{i}= & \bigl(\pm J\hat{\boldsymbol{\mathrm{k}}}_{i}-\omega_{\mathrm{a}}\boldsymbol{\hat{\mathrm{f}}}_{i}\bigr)\times\boldsymbol{e}_{x}\pm Q\hat{S}_{z}\boldsymbol{e}_{y}-\gamma_{\mathrm{a}}\boldsymbol{\hat{\mathrm{f}}}_{i}+\boldsymbol{\hat{F}}_{i\mathrm{a}}\label{eq:J dynamics-1}\\
\partial_{t}\hat{\boldsymbol{\mathrm{k}}}_{i}= & \bigl(\pm J\boldsymbol{\hat{\mathrm{f}}}_{i}-\omega_{\mathrm{b}}\hat{\boldsymbol{\mathrm{k}}}_{i}\bigr)\times\boldsymbol{e}_{x}-\gamma_{\mathrm{b}}\hat{\boldsymbol{\mathrm{k}}}_{i}+\boldsymbol{\hat{F}}_{i\mathrm{b}}\label{eq:K dynamics-1}
\end{align}
Here $\gamma_{\text{L}}$ denotes the attenuation per-unit-length
of the probe (including the absorption by the alkali atoms), $\gamma_{\mathrm{a}}$
denotes the total decoherence rate of the alkali spins in the presence
of the probe, and $\gamma_{\mathrm{b}}$ denotes the slow relaxation
of the noble-gas spins \cite{Polzik-two-cell-theory}. The vacuum
noise operators $\boldsymbol{\hat{F}}_{\mathrm{L}}$ , $\boldsymbol{\hat{F}}_{i\mathrm{a}}$,
and $\boldsymbol{\hat{F}}_{i\mathrm{b}}$ are associated with these
decays \cite{Firstenberg-Weak-collisions,Polzik-RMP-2010}. The spin-exchange
interaction allows for coherent state-exchange between the alkali
and noble-gas spins within each of the two cells independently at
a rate $J=g\sqrt{M_{\mathrm{a}}M_{\mathrm{b}}}/(AL)$. The coherent
spin-exchange rate coefficient is $g$, with $g=4.9\times10^{-15}\,\text{\ensuremath{\unit{cm^{3}}}}\text{s}^{-1}$
for a $\text{K\ensuremath{-^{3}}He}$ mixture or $g=1.9\times10^{-13}\,\text{\ensuremath{\unit{cm^{3}}}}\text{s}^{-1}$
for $^{87}\text{Rb\ensuremath{-^{129}}Xe}$ \cite{Firstenberg-Weak-collisions,Happer-Book}.
At the same time, the polarization state of the probe is altered by
both ensembles together: $\hat{S}_{y}$ depends on the non-local spin
operator $\hat{\mathrm{f}}_{1z}+\hat{\mathrm{f}}_{2z}$, and $\hat{S}_{z}$
exerts a common back-action light-shift on the two cells. The optical
coupling rate $Q=$$(a/T)\sqrt{M_{\mathrm{a}}M_{\mathrm{L}}}$ depends
on $a=2r_{\mathrm{e}}cf/[A\delta_{\mathrm{e}}(2I+1)]$, where $r_{\mathrm{e}}=2.8\times10^{-17}$~cm
is the classical electron radius, $f\leq1$ is the oscillator strength
of the atomic transition, and $\delta_{\mathrm{e}}$ is the detuning
of the laser from the optical transition. Also note that the operators
in Eqs.~(\ref{eq: S stokes dynamics-1}-\ref{eq:K dynamics-1}) satisfy
the commutation relations $[\hat{\mathrm{f}}_{iy},\hat{\mathrm{f}}_{jz}]=\pm i\delta_{ij}$
for the alkali, $\bigl[\hat{S}_{y}(z'),\hat{S}_{z}(z'')\bigr]=icT\delta(z'-z'')$
for the light and $[\hat{\mathrm{k}}_{iy},\hat{\mathrm{k}}_{jz}]=\pm i\delta_{ij}$
for the noble-gas spins.

To simplify Eqs.~(\ref{eq: S stokes dynamics-1}-\ref{eq:K dynamics-1}),
we transform the system to the rotating frame of the noble-gas spins
and describe the adiabatic following of the alkali in the limit of
large magnetic field limit (the off-resonance regime) $\Delta\gg\gamma_{\mathrm{a}},J,Q$.
The formal transformation of the collective spin operators in each
cell to the rotating frame is given by $\boldsymbol{\hat{f}}_{i}'=R_{x}(\omega_{\mathrm{b}})\boldsymbol{\hat{f}}_{i}$
and $\boldsymbol{\hat{k}}_{i}'=R_{x}(\omega_{\mathrm{b}})\boldsymbol{\hat{k}}{}_{i}$,
using the standard rotation matrix
\begin{equation}
R_{x}(\omega_{\mathrm{b}})=\left(\begin{array}{ccc}
1 & 0 & 0\\
0 & \cos(\omega_{\mathrm{b}}t) & \sin(\omega_{\mathrm{b}}t)\\
0 & -\sin(\omega_{\mathrm{b}}t) & \cos(\omega_{\mathrm{b}}t)
\end{array}\right).
\end{equation}
The operators $\boldsymbol{\hat{f}}'$ and $\boldsymbol{\hat{k}}'$
are the stationary spin components of the alkali and noble-gas spins,
respectively. The dynamics of the $y,z$ components of the alkali
spins in the rotating frame is then given by 
\begin{equation}
\partial_{t}\boldsymbol{\hat{\mathrm{f}}}_{i}'=\bigl(\pm J\hat{\boldsymbol{\mathrm{k}}}_{i}'-\Delta\boldsymbol{\hat{\mathrm{f}}}_{i}'\bigr)\times\boldsymbol{e}_{x}\pm Q\hat{S}_{z}\boldsymbol{e}_{y}(t)-\gamma_{\mathrm{a}}\boldsymbol{\hat{\mathrm{f}}}_{i}'+\boldsymbol{\hat{F}}_{i\mathrm{a}}'.
\end{equation}
We are interested in the slow, adiabatic dynamics of $\boldsymbol{\hat{\mathrm{f}}}_{i}'$,
which naturally oscillates at a rate $\Delta$. The leading order
of the dynamics is thus determined by considering the instantaneous
steady state $\partial_{t}\boldsymbol{\hat{\mathrm{f}}}_{i}'=0$,
which yields the linear relation
\begin{equation}
\boldsymbol{\hat{\mathrm{f}}}_{i}'=\frac{\pm1}{\sqrt{\Delta^{2}+\gamma_{\mathrm{a}}^{2}}}R_{x}(\psi)\cdot\left(Q\hat{S}_{z}\boldsymbol{e}(t)+J\boldsymbol{\hat{\mathrm{k}}}_{i}'+\boldsymbol{\hat{F}}_{i\mathrm{a}}'\right).\label{eq:alkali_adiabatic_spins}
\end{equation}
Here $\boldsymbol{e}(t)=\sin(\omega_{\mathrm{b}}t)\boldsymbol{e_{y}}+\cos(\omega_{\mathrm{b}}t)\boldsymbol{e_{z}}$
is the optical axis in the rotating frame, and we define $\cos\psi\equiv\Delta/\sqrt{\Delta^{2}+\gamma_{\mathrm{a}}^{2}}$
and $\sin\psi\equiv\gamma_{\mathrm{a}}/\sqrt{\Delta^{2}+\gamma_{\mathrm{a}}^{2}}$.
Eq.~(\ref{eq:alkali_adiabatic_spins}) describes the slow temporal
dependence of the alkali spin operators on the noble-gas spins via
spin-exchange, on the light circular polarization via back-action
noise, and on the infiltrated vacuum white noise associated with the
decay rate $\gamma_{\mathrm{a}}$. The noise terms are given by $\boldsymbol{\hat{F}}_{ia}'=R_{x}(\omega_{\mathrm{b}})\boldsymbol{\hat{F}}_{ia}$,
which are statistically equivalent to $\boldsymbol{\hat{F}}_{ia}$.
In the off-resonance regime $\Delta\gg\gamma_{\mathrm{a}}$, we obtain
$\psi\ll1$, such that the leading term in Eq.~(\ref{eq:alkali_adiabatic_spins})
is free of decay and noise, yielding the simple form of Eq.~(3) in
the main text (note that in the main text, we dropped the prime notation
for brevity). 

We now substitute Eq.~(\ref{eq:alkali_adiabatic_spins}) in Eqs.~(\ref{eq: S stokes dynamics-1})
and (\ref{eq:K dynamics-1}) to obtain the dynamics of $\boldsymbol{\hat{S}}$
as a function of $\hat{\boldsymbol{\mathrm{k}}}_{i}'$

\begin{align}
\partial_{z}\boldsymbol{\hat{S}} & =\frac{TQ}{L}\left(\cos(\omega_{\mathrm{b}}t)\bigl(\hat{\mathrm{f}}_{1z}'+\hat{\mathrm{f}}_{2z}'\bigr)+\sin(\omega_{\mathrm{b}}t)\bigl(\hat{\mathrm{f}}_{1y}'+\hat{\mathrm{f}}_{2y}'\bigr)\right)\boldsymbol{e}_{y}-\gamma_{\text{L}}\boldsymbol{\hat{S}}+\boldsymbol{\hat{F}}_{\mathrm{L}}\label{eq:S_adiabatic_intro}\\
= & \frac{1}{L}\frac{TQ}{\sqrt{\Delta^{2}+\gamma_{\mathrm{a}}^{2}}}\sum_{i=1}^{2}\left(-1\right)^{i+1}\left[(Q\hat{S}_{z}+J\hat{\mathrm{k}}_{iz}+\hat{F}_{i\mathrm{a,y}})\cos\psi-(J\hat{\mathrm{k}}_{iy}+\hat{F}_{i\mathrm{a,z}})\sin\psi\right]\boldsymbol{e}_{y}-\gamma_{\text{L}}\boldsymbol{\hat{S}}+\boldsymbol{\hat{F}}_{\mathrm{L}}\label{eq:S_adiabatic_step1}\\
= & \frac{1}{L}\frac{TQ}{\sqrt{\Delta^{2}+\gamma_{\mathrm{a}}^{2}}}\left[J\left(\cos\psi(\hat{\mathrm{k}}_{1z}-\hat{\mathrm{k}}_{2z})-\sin\psi(\hat{\mathrm{k}}_{1y}-\hat{\mathrm{k}}_{2y})\right)+\sum_{i=1}^{2}\left(-1\right)^{i+1}\left(\cos\psi\hat{F}_{i\mathrm{a,y}}+\sin\psi\hat{F}_{i\mathrm{a,z}}\right)\right]\boldsymbol{e}_{y}-\gamma_{\text{L}}\boldsymbol{\hat{S}}+\boldsymbol{\hat{F}}_{\mathrm{L}}\label{eq:S_adiabatic_step2}\\
= & \frac{1}{L}\frac{TQ}{\sqrt{\Delta^{2}+\gamma_{\mathrm{a}}^{2}}}\left[J\left(\sin(\omega_{\mathrm{b}}t-\psi)(\hat{\mathrm{k}}_{1y}'-\hat{\mathrm{k}}_{2y}')+\cos(\omega_{\mathrm{b}}t-\psi)(\hat{\mathrm{k}}_{1z}'-\hat{\mathrm{k}}_{2z}')\right)+\sqrt{2}\hat{F}_{\mathrm{a,y}}\right]\boldsymbol{e}_{y}-\gamma_{\text{L}}\boldsymbol{\hat{S}}+\boldsymbol{\hat{F}}_{\mathrm{L}}\label{eq:S_adiabatic_step3}
\end{align}
We first note, when moving from Eq.~(\ref{eq:S_adiabatic_step1})
to Eq.~(\ref{eq:S_adiabatic_step2}), that the dependence of $\hat{S}_{y}$
on $\hat{S}_{z}$- also known as polarization self-rotation \cite{Rochester-2001}-
is canceled by using the double cell configuration with opposite alkali
spin polarization. In Eq.~(\ref{eq:S_adiabatic_step3}), we use the
identity $R_{x}(\omega_{\mathrm{b}}t)R_{x}(-\psi)=R_{x}(\omega_{\mathrm{b}}t-\psi)$
and identify the noise process $\hat{F}_{\mathrm{a,y}}\equiv\sum_{i=1}^{2}\left(-1\right)^{i+1}(\cos\psi\hat{F}_{i\mathrm{a,y}}+\sin\psi\hat{F}_{i\mathrm{a,z}})/\sqrt{2}$,
which has identical statistics as $\hat{F}_{1\mathrm{a,y}}$ and $\hat{F}_{1\mathrm{a,z}}$.
For negligible loss mechanisms $\psi,\gamma_{\mathrm{L}}L\ll1,$ Eq.~(\ref{eq:S_adiabatic_step3})
provides Eq.~(5) in the main text. We also note that in the absence
of loss, $S_{z}$ remains constant throughout the cell.

Similarly, we derive the equations for the noble-gas spins in each
cell in the rotating frame
\begin{align}
\partial_{t}\hat{\boldsymbol{\mathrm{k}}}_{i}' & =\pm JR_{x}(\tfrac{\pi}{2})\boldsymbol{\hat{\mathrm{f}}}_{i}'-\gamma_{\mathrm{b}}\hat{\boldsymbol{\mathrm{k}}}_{i}'+\boldsymbol{\hat{F}}_{i\mathrm{b}}\\
= & \frac{J}{\sqrt{\Delta^{2}+\gamma_{\mathrm{a}}^{2}}}R_{x}(\psi)R_{x}(\tfrac{\pi}{2})\cdot\left(Q\hat{S}_{z}\boldsymbol{e}(t)+J\boldsymbol{\hat{\mathrm{k}}}_{i}'+\boldsymbol{\hat{F}}_{i\mathrm{a}}'\right)-\gamma_{\mathrm{b}}\hat{\boldsymbol{\mathrm{k}}}_{i}'+\boldsymbol{\hat{F}}_{i\mathrm{b}}\\
= & \frac{JQ}{\sqrt{\Delta^{2}+\gamma_{\mathrm{a}}^{2}}}\hat{S}_{z}\left(\cos(\omega_{\mathrm{b}}t+\psi)\boldsymbol{e}_{y}-\cos(\omega_{\mathrm{b}}t+\psi)\boldsymbol{e}_{z}\right)+\delta\omega_{\mathrm{b}}\hat{\boldsymbol{\mathrm{k}}}_{i}'\times\boldsymbol{e}_{x}-\Gamma_{\mathrm{b}}\hat{\boldsymbol{\mathrm{k}}}_{i}'+\boldsymbol{\hat{F}}_{i\mathrm{b}}'.
\end{align}
The first term describes the back-action of the probe circular polarization
on the noble-gas spins, mediated via its effect on the alkali. Interestingly,
the back-action is the same in both cells, being polarized in opposite
orientations. The second term, with $\delta\omega_{\mathrm{b}}=\Delta J^{2}/(\Delta^{2}+\gamma_{\mathrm{a}}^{2})$,
describes a small spin-exchange induced shift, typically satisfying
$\delta\omega_{\mathrm{b}}\ll\omega_{\mathrm{b}}$. This small shift
can be taken into account by the simple transformation $\omega_{\mathrm{b}}\rightarrow\omega_{\mathrm{b}}-\delta\omega_{\mathrm{b}}$
in the equations. The third term describes the decoherence of the
noble-gas spins with the total rate $\Gamma_{\mathrm{b}}=\gamma_{\mathrm{b}}+\gamma_{\mathrm{a}}J^{2}/(\Delta^{2}+\gamma_{\mathrm{a}}^{2})$,
which includes the alkali-induced relaxation as described in the main
text. $\boldsymbol{\hat{F}}_{i\mathrm{b}}'$ denotes a quantum white
noise operator associated with the decay rate $\Gamma_{\mathrm{b}}$.

We now consider the dynamics of the nonlocal noble-gas spin operators.
For the difference of the spin operators $\hat{\boldsymbol{\mathrm{k}}}_{1}'-\hat{\boldsymbol{\mathrm{k}}}_{2}'$,
we find
\begin{equation}
\partial_{t}(\hat{\boldsymbol{\mathrm{k}}}_{1}'-\hat{\boldsymbol{\mathrm{k}}}_{2}')=-\Gamma_{\mathrm{b}}(\hat{\boldsymbol{\mathrm{k}}}_{1}'-\hat{\boldsymbol{\mathrm{k}}}_{2}')+\sqrt{2}\boldsymbol{\hat{F}}_{\mathrm{b}-},\label{eq:spin difference}
\end{equation}
where we define $\boldsymbol{\hat{F}}_{\mathrm{b}\pm}\equiv(\boldsymbol{\hat{F}}_{1b}'\pm\boldsymbol{\hat{F}}_{2\mathrm{b}}')\sqrt{2}$.
Here the common probe back-action is canceled, in accordance with
the requirements of the EPR entanglement criterion {[}cf. Eq.~(1)
in the main text{]}. For low noble-gas losses $\Gamma_{\mathrm{b}}T\ll1$,
Eq.~(\ref{eq:spin difference}) becomes Eq.~(4) in the main text.
The sum of spin operators, is significantly affected by the probe
back-action, given by
\begin{equation}
\partial_{t}(\hat{\boldsymbol{\mathrm{k}}}_{1}'+\hat{\boldsymbol{\mathrm{k}}}_{2}')=\frac{2JQ}{\sqrt{\Delta^{2}+\gamma_{\mathrm{a}}^{2}}}\hat{S}_{z}\left(\cos(\omega_{\mathrm{b}}t+\psi)\boldsymbol{e}_{y}-\cos(\omega_{\mathrm{b}}t+\psi)\boldsymbol{e}_{z}\right)-\Gamma_{\mathrm{b}}(\hat{\boldsymbol{\mathrm{k}}}_{1}'+\hat{\boldsymbol{\mathrm{k}}}_{2}')+\sqrt{2}\boldsymbol{\hat{F}}_{\mathrm{b+}}.\label{eq:spin sum}
\end{equation}

Before solving Eqs.~(\ref{eq:S_adiabatic_step3}) and (\ref{eq:spin difference}-\ref{eq:spin sum}),
we renormalize the operators and transform them to the canonical form
\cite{Polzik-RMP-2010}. Using the definitions in the main text, we
find two independent sets of harmonic oscillator operators, which
we here explicitly identify. The first set describes the in-phase
component of the light polarization operators
\begin{align}
\hat{x}_{\text{L},y}\left(z\right) & =\frac{\sqrt{2}}{T}\int_{0}^{T}\hat{S}_{y}\left(z\right)\sin(\omega_{\mathrm{b}}t)dt,\label{eq:X_ly}\\
\hat{p}_{\text{L},y}\left(z\right) & =\frac{\sqrt{2}}{T}\int_{0}^{T}\hat{S}_{z}\left(z\right)\sin(\omega_{\mathrm{b}}t)dt,\label{eq:P_ly}
\end{align}
coupled to the spin operators 
\begin{align}
\hat{x}_{\text{b},y}\left(t\right) & =-\frac{1}{\sqrt{2}}(\hat{\mathrm{k}}_{1z}'+\hat{\mathrm{k}}_{2z}'),\label{eq:X_by}\\
\hat{p}_{\text{b},y}\left(t\right) & =\frac{1}{\sqrt{2}}(\hat{\mathrm{k}}_{1y}'-\hat{\mathrm{k}}_{2y}').\label{eq:P_by}
\end{align}
The second set is given by the operators 

\begin{align}
\hat{x}_{\text{L},z}\left(z\right) & =\frac{\sqrt{2}}{T}\int_{0}^{T}\hat{S}_{y}\left(z\right)\cos(\omega_{\mathrm{b}}t)dt\label{eq:X_lz}\\
\hat{p}_{\text{L},z}\left(z\right) & =\frac{\sqrt{2}}{T}\int_{0}^{T}\hat{S}_{z}\left(z\right)\cos(\omega_{\mathrm{b}}t)dt\label{eq:P_lz}\\
\hat{x}_{\text{b},z}\left(t\right) & =\frac{1}{\sqrt{2}}(\hat{\mathrm{k}}_{1y}'+\hat{\mathrm{k}}_{2y}')\label{eq:X_bz}\\
\hat{p}_{\text{b},z}\left(t\right) & =\frac{1}{\sqrt{2}}(\hat{\mathrm{k}}_{1z}'-\hat{\mathrm{k}}_{2z}').\label{eq:P_bz}
\end{align}
The light vector operators consist of two sets of standard Harmonic
oscillator operators satisfying $\bigl[\hat{x}_{\text{L},\alpha},\hat{p}_{\text{L},\beta}\bigr]=i\delta_{\alpha\beta}$
for $\alpha,\beta\in\left\{ y,z\right\} $. Similarly, the atomic
operators satisfy $\bigl[\hat{x}_{\text{b},\alpha},\hat{p}_{\text{b},\beta}\bigr]=i\delta_{\alpha\beta}$. 

We first derive the leading terms in the input and output relations
of the system, absent relaxation and noise. Substituting the definitions
(\ref{eq:X_ly}-\ref{eq:P_bz}) into Eq.~(\ref{eq:S_adiabatic_step3})
and spatially integrating along the beam path yield the light response
to the emerging Faraday interaction with noble-gas spins 
\begin{align}
\boldsymbol{\hat{x}}_{\text{L}}^{\text{out}} & =\boldsymbol{\hat{x}}_{\text{L}}^{\text{in}}+\kappa\boldsymbol{\hat{p}}_{\text{b}}^{\text{in}},\label{eq:Xl_in_out}\\
\boldsymbol{\hat{p}}_{\text{L}}^{\text{out}} & =\boldsymbol{\hat{p}}_{\text{L}}^{\text{in}},
\end{align}
where 
\begin{equation}
\kappa=\frac{JQT}{\sqrt{\Delta^{2}+\gamma_{\mathrm{a}}^{2}}}\label{eq:kappa}
\end{equation}
is the unitless optical coupling strength. Similarly, substituting
the definitions (\ref{eq:X_ly}-\ref{eq:P_bz}) in Eqs.~(\ref{eq:spin difference}-\ref{eq:spin sum})
and temporally integrating for the pulse duration $T$ yield the atomic
evolution of the spins by the emerging Faraday interaction with noble-gas
spins 
\begin{align}
\boldsymbol{\hat{x}}_{\text{b}}^{\text{out}} & =\boldsymbol{\hat{x}}_{\text{b}}^{\text{in}}+\kappa\boldsymbol{\hat{p}}_{\text{L}}^{\text{in}},\\
\boldsymbol{\hat{p}}_{\text{b}}^{\text{out}} & =\boldsymbol{\hat{p}}_{\text{b}}^{\text{in}}.\label{eq:Pb_in_out}
\end{align}
Equations.~(\ref{eq:Xl_in_out}) and (\ref{eq:Pb_in_out}) give Eqs.~(6)
in the main text. In the presence of noise and relaxations, the modified
input-output relations are given by 
\begin{align}
\boldsymbol{\hat{x}}_{\text{L}}^{\text{out}} & =\sqrt{1-\epsilon}\bigl(\boldsymbol{\hat{x}}_{\text{L}}^{\text{in}}+\kappa\boldsymbol{\hat{p}}_{\text{b}}^{\text{in}}+\kappa\sqrt{\varrho}\boldsymbol{\hat{w}}_{0}\bigr)+\sqrt{\epsilon}\boldsymbol{\hat{w}}_{1}\nonumber \\
\boldsymbol{\hat{p}}_{\text{L}}^{\text{out}} & =\sqrt{1-\epsilon}\boldsymbol{\hat{p}}_{\text{L}}^{\text{in}}+\sqrt{\epsilon}\boldsymbol{\hat{w}}_{2}\nonumber \\
\boldsymbol{\hat{x}}_{\text{b}}^{\text{out}} & =\sqrt{1-\eta}\bigl(\boldsymbol{\hat{x}}_{\text{b}}^{\text{in}}+\kappa\boldsymbol{\hat{p}}_{\text{L}}^{\text{in}}\bigr)+\sqrt{\eta}\boldsymbol{\hat{w}}_{3}\\
\boldsymbol{\hat{p}}_{\text{b}}^{\text{out}} & =\sqrt{1-\eta}\boldsymbol{\hat{p}}_{\text{b}}^{\text{in}}+\sqrt{\eta}\boldsymbol{\hat{w}}_{4}.\nonumber 
\end{align}
Here we identify $\boldsymbol{\hat{w}}_{n}$ ($0\leq n\leq4)$ as
standard vacuum-noise operators which correspond to normalized quantum-Weiner
processes, satisfying $\langle\boldsymbol{\hat{w}}_{n}\rangle=0$
and $\langle\hat{w}_{m\alpha}\hat{w}_{n\beta}\rangle=\frac{1}{2}\delta_{mn}\delta_{\alpha\beta}$
for $\alpha,\beta\in\left\{ y,z\right\} $ and $0\leq m,n\leq4$ \cite{Polzik-RMP-2010}. 

To estimate the attainable degree of squeezing and choose the optimal
feedback pulse, we calculate the variance of the atomic spins after
the feedback $\text{var}(p_{A,i}^{\text{out}}+Gx_{L,z}^{\text{out}})$
and find that it attains a minimal value of 
\begin{equation}
\text{var}(p_{A,i}^{\text{out}}+Gx_{L,z}^{\text{out}})=\frac{1}{2}\left(\frac{\kappa^{2}\left(1-\epsilon\right)\left(\eta+\varrho\right)+1}{\kappa^{2}\left(1-\epsilon\right)\left(1+\varrho\right)+1}\right)
\end{equation}
for the feedback proportionality constant 
\begin{equation}
G=-\frac{\kappa\sqrt{1-\epsilon}\sqrt{1-\eta}}{\left(1+\kappa^{2}\left(\varrho+1\right)\left(1-\epsilon\right)\right)}.
\end{equation}

\section{Entanglement lifetime}
\begin{figure}[t]
\begin{centering}
\includegraphics[clip,width=15cm]{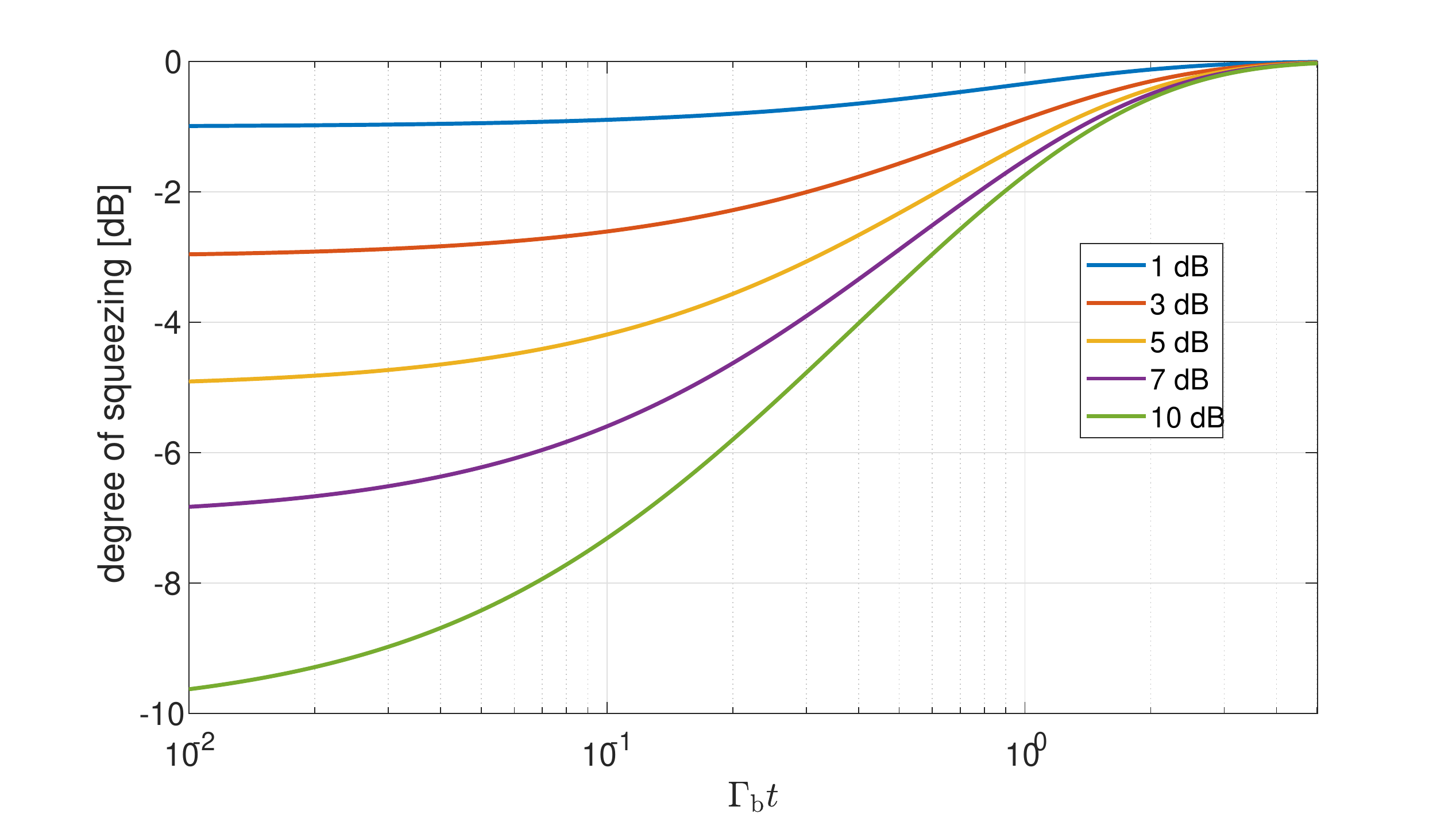}
\par\end{centering}
\centering{}\caption{Degree of squeezing for vacuum-squeezed states with initial values
of squeezing of $1,3,5,7$ and $10\,\text{dB}$. While in a linear
scale the variance decays at a constant rate $2\Gamma_{\mathrm{b}}$,
representation in a logarithmic (dB) scale gives the time dependence
shown \label{fig: squeezing_degree_time}}
\end{figure}

\setcounter{equation}{32}
In this section, we show that the variance of the squeezed quadrature
of the two ensembles decays at the rate $2\Gamma_{\mathrm{b}}$. In
our case, the two independently squeezed quadratures are $\hat{p}_{\text{b},y}\left(t\right)$
and $\hat{p}_{\text{b},z}\left(t\right)$, which according to Eq.~(\ref{eq:spin difference})
satisfy the dynamics

\begin{equation}
\partial_{t}\boldsymbol{\hat{p}}_{\text{b}}=-\Gamma_{\mathrm{b}}\boldsymbol{\hat{p}}_{\text{b}}+\boldsymbol{\hat{F}}_{\mathrm{b}-}\label{eq:spin difference-1}
\end{equation}
where $\langle\boldsymbol{\hat{F}}_{\mathrm{b}-}\rangle=0,$ and 
\begin{equation}
\langle\hat{F}_{\mathrm{b}-,\alpha}(t-t')\hat{F}_{\mathrm{b}-,\beta}(t-t'')\rangle=\Gamma_{\mathrm{b}}\delta(t''-t')
\end{equation}
for $\alpha,\beta\in\left\{ y,z\right\} $. Integration of Eq.~(\ref{eq:spin difference-1})
yields
\begin{equation}
\boldsymbol{\hat{p}}_{\text{b}}\left(t\right)=e^{-\Gamma_{\mathrm{b}}t}\boldsymbol{\hat{p}}_{\text{b}}\left(0\right)+\int_{0}^{t}e^{-\Gamma_{\mathrm{b}}\left(t-t'\right)}\boldsymbol{\hat{p}}_{\text{b}}\left(t'\right)dt',
\end{equation}
where the second integral represents a standard stochastic integration.
We first note that the initial vacuum-squeezed state is not displaced
yielding $\langle\hat{p}_{\text{b}\alpha}\left(0\right)\rangle=0$
and this $\langle\hat{p}_{\text{b}\alpha}\left(t\right)\rangle=0$.
To calculate the variance as a function of time we first find
\begin{equation}
\hat{p}_{\text{b}\alpha}^{2}\left(t\right)=e^{-2\Gamma_{\mathrm{b}}t}\hat{p}_{\text{b}\alpha}^{2}\left(0\right)+\int_{0}^{t}e^{-\Gamma_{\mathrm{b}}\left(2t-t'\right)}\left\{ \hat{p}_{\text{b}\alpha}\left(0\right),\hat{F}_{\mathrm{b}-,\alpha}\left(t'\right)\right\} dt'+\int_{0}^{t}\int_{0}^{t}e^{-\Gamma_{\mathrm{b}}\left(2t-t'-t''\right)}\hat{F}_{\mathrm{b}-,\alpha}\left(t'\right)\hat{F}_{\mathrm{b}-,\alpha}\left(t''\right)dt'dt''
\end{equation}
where $\left\{ \cdot\right\} $ denotes the anti commutator. We can
than calculate the variance by
\begin{align}
\text{var}\left(\hat{p}_{\text{b}\alpha}\left(t\right)\right) & =e^{-2\Gamma_{\mathrm{b}}t}\langle\hat{p}_{\text{b}\alpha}^{2}\left(0\right)\rangle+\int_{0}^{t}\int_{0}^{t}e^{-\Gamma_{\mathrm{b}}\left(2t-t'-t''\right)}\langle\hat{F}_{\mathrm{b}-,\alpha}\left(t'\right)\hat{F}_{\mathrm{b}-,\alpha}\left(t''\right)\rangle dt'dt''\\
= & e^{-2\Gamma_{\mathrm{b}}t}\langle\hat{p}_{\text{b}\alpha}^{2}\left(0\right)\rangle+\Gamma_{\mathrm{b}}\int_{0}^{t}e^{-2\Gamma_{\mathrm{b}}\left(t-t'\right)}dt'\nonumber \\
= & \frac{1}{2}+\left(\langle\hat{p}_{\text{b}\alpha}^{2}\left(0\right)\rangle-\frac{1}{2}\right)e^{-2\Gamma_{\mathrm{b}}t}.
\end{align}
Therefore, the variance of a squeezed state initially with $\langle\hat{p}_{\text{b}\alpha}^{2}\left(0\right)\rangle<\frac{1}{2}$
would decay at the rate $2\Gamma_{\mathrm{b}}$(i.e. twice the individual
decoherence rate), whereas $\text{std}(\hat{p}_{\text{b}\alpha})$
decays at a rate $\Gamma_{\mathrm{b}}$. Note that if the degree of
squeezing is represented in dB scale via the definition $10\log_{10}\left(2\text{var}\left(\hat{p}_{\text{b}\alpha}\left(t\right)\right)\right)$,
then the decay would seem faster for higher degree of squeezing as
demonstrated in Fig.~\ref{fig: squeezing_degree_time} for different
initial squeezing degrees. 

\section{Additional experimental configurations}

The entanglement generation scheme can be realized with various alkali
and noble-gas mixtures within a large range of experimental parameters.
Here we consider two additional configurations using two cylindrical
cells of length $L=5$ cm and cross-section $A=2\,\text{mm}^{2}$.
First, we consider entanglement of two $^{129}\text{Xe}$ ensembles
each comprising gaseous mixture of 5~Torr $^{129}\text{Xe}$, 50~Torr
$\text{N}_{2}$, 650~Torr $\text{Ne}$, and a droplet of $^{87}\mathrm{Rb}$
at $175^{\circ}\text{C}$. The $\text{N}_{2}$ serves for quenching,
and the Ne serves as a buffer gas that also acts to break XeRb molecules.
Optical pumping of the Rb spins with $R_{\text{op}}=1.6\gamma_{\mathrm{a}}$
yields $P_{\mathrm{a}}=0.62$ {[}with $q(3/2,P_{\mathrm{a}})=1.22]$.
The $^{129}\text{Xe}$ is hyperpolarized via SEOP to $P_{\mathrm{b}}=0.46$,
and its expected decoherence time $\gamma_{\mathrm{b}}^{-1}\approx5\,\mathrm{sec}$
is dominated by collisions with alkali atoms \cite{Happer-Book,Happer-1984}.
Magnetic field of order $B_{1}\approx20\,\text{mG}$ is applied. A
250-mW probe detuned $980\,\text{GHz}$ from the optical line is measured
via homodyne detection for $T=200\,\text{msec}$. These yield $\kappa=1.8$,
$\epsilon=0.28$, $\eta=0.22$, $\varrho=0.17$, and could generate
3~dB of two-mode squeezing $(\xi=0.34)$. After the generation of
entanglement, the optical pumping is turned off, the alkali depolarizes
quickly, and the entanglement between the noble-gas ensembles can
persist for several seconds.

Second, we consider entanglement of two $^{3}\text{He}$ ensembles
using a gaseous mixture of 50~Torr $^{3}\text{He}$, 100~Torr $\text{N}_{2}$,
and a droplet of K at $250^{\circ}\text{C}$. Here $R_{\text{op}}=1.1\gamma_{\mathrm{a}}$
yields $P_{\mathrm{a}}=0.52$ {[}with $q(3/2,P_{\mathrm{a}})=1.28${]}
and $P_{\mathrm{b}}=0.48$, assuming $\gamma_{\mathrm{b}}^{-1}=50\,\text{hour}$.
The 400-mW probe is detuned $1\,\text{THz}$ from the optical line,
and $B_{1}\approx70\,\text{mG}$. Homodyne detection for $T=2\,\text{minutes}$
yields $\kappa=2.9$, $\epsilon=0.3$, $\eta=0.12$ and $\varrho=0.02$
, generating almost 6~dB of two-mode squeezing $(\xi=0.68)$, which
could live for hours.

\section{dependence of the interaction strength on the optical depth}
\setcounter{equation}{38}

In the limit $\epsilon,\eta,\varrho\ll1$, the attainable entanglement
is governed by $\kappa$ which characterizes the polarization rotation
induced by the (noble-gas) spin-noise and imprinted on the probe signal
with respect to the photon shot-noise A similar parameter, $\tilde{\kappa}$,
governs the attainable entanglement between alkali-spin ensembles
(in the absence of noble-gas), where it is known that $\tilde{\kappa}^{2}=2\gamma_{\mathrm{a}}Td$
\cite{Polzik-RMP-2010}. There, one assumes that the alkali relaxation
$\gamma_{\mathrm{a}}$ is dominated by scattering of probe photons,
and one identifies the resonant optical-dep\textcolor{black}{th $d$
as the primary resource for entanglement. Similarly in our scheme,
when the relaxation inherited from the alkali atoms $(J^{2}/\Delta^{2})\gamma_{\mathrm{a}}$
dominates $\Gamma_{\mathrm{b}}$, the relation $\kappa^{2}=2\Gamma_{\mathrm{b}}Td$
also holds} as we now derive in the off-resonance regime: 
\begin{align}
\kappa^{2} & =\frac{J^{2}Q^{2}T^{2}}{\Delta^{2}+\gamma_{\mathrm{a}}^{2}}=T^{2}\frac{M_{\mathrm{a}}M_{\mathrm{L}}a^{2}}{T^{2}}\frac{J^{2}}{\Delta^{2}+\gamma_{\mathrm{a}}^{2}}=a^{2}M_{\mathrm{a}}M_{\mathrm{L}}\frac{J^{2}}{\Delta^{2}+\gamma_{\mathrm{a}}^{2}}\nonumber \\
= & \left(\frac{\Gamma_{\mathrm{e}}^{2}\sigma^{2}}{A^{2}\delta_{\mathrm{e}}^{2}(2I+1)^{2}}\right)N_{\mathrm{a}}(I+1/2)M_{\mathrm{L}}\frac{J^{2}}{\Delta^{2}+\gamma_{\mathrm{a}}^{2}}=T\left(\frac{M_{\mathrm{L}}\sigma}{T(2I+1)A}\frac{\Gamma_{\mathrm{e}}^{2}}{\delta_{\mathrm{e}}^{2}}\right)\frac{n_{\mathrm{a}}\sigma L}{2}\frac{J^{2}}{\Delta^{2}+\gamma_{\mathrm{a}}^{2}}\nonumber \\
= & 2\left(\frac{\gamma_{\text{absp}}}{\gamma_{\mathrm{a}}}\frac{J^{2}\gamma_{\mathrm{a}}}{\Delta^{2}+\gamma_{\mathrm{a}}^{2}}\right)Td
\end{align}
where we use the on-resonance cross-section of a hot vapor with homogeneous
broadening as $\sigma=2r_{\mathrm{e}}cf/\Gamma_{\mathrm{e}}$. We
also identify the number of alkali atoms $N_{a}=n_{a}AL$ and the
on-resonance pumping rate $R_{a}=M_{\mathrm{L}}\sigma/(T(2I+1)A)$,
including the slowing-down factor of polarized ensembles. The off-resonance
absorption rate is then $\gamma_{\text{absp}}=R_{a}\Gamma_{\mathrm{e}}^{2}/(4\delta_{\mathrm{e}}^{2})$.
The optical-depth of a single cell is $d=n_{\mathrm{a}}\sigma L$.

Assuming that the alkali relaxation is dominated by absorption of
probe photons, one can set $\gamma_{\mathrm{a}}\approx\gamma_{\text{absp}}$.
Further assuming that the dominant relaxation of the noble-gas is
due to its coupling to the alkali, $\Gamma_{\mathrm{b}}\approx J^{2}\gamma_{\mathrm{a}}/(\Delta^{2}+\gamma_{\mathrm{a}}^{2})$,
yields the relation 
\begin{equation}
\kappa^{2}\approx2\Gamma_{\mathrm{b}}Td.
\end{equation}
\textcolor{black}{Th}erefore, the optical-depth of the alkali ensembles
remains the primary resource also for entangling noble-gas spins.
\end{document}